# From Players to Participants: Citizen Science and Video Games to Understand Cognition

Syrine Salouhou[1], Edgar Dubourg[2,3], Maxwell Scott-Slade[5], Hugo Spiers[4], Antoine Coutrot[1]

1. CNRS, Ecole Centrale de Lyon, INSA Lyon, Universite Claude Bernard Lyon 1, Université Lumière Lyon 2, LIRIS, UMR5205, Lyon, France
2. Institut Curie, PSL Research University, 26 rue d'Ulm, 75248, Paris Cedex 05, France
3. Institut Jean Nicod, Département d'études cognitives, Ecole normale supérieure, Université PSL, EHESS, CNRS, 75005 Paris, France
4. Institute of Behavioural Neuroscience, Department of Experimental Psychology, University College London, UK
5. Glitchers Ltd., Edinburgh, Scotland, United Kingdom.

## Abstract

Citizen science is transforming how cognitive scientists study the human mind, and video games are at the heart of this shift. By embedding experimental tasks into engaging, game-like experiences, researchers can reach large, diverse populations while collecting rich behavioral data outside the lab. In this review, we explore how citizen science video games bridge the gap between players and participants, turning entertainment into large-scale cognitive research. Drawing on recent projects such as *Sea Hero Quest* and *The Music Lab*, we outline the key benefits of this approach: scalability, ecological validity, and public engagement. We also examine the challenges of designing games that are scientifically rigorous, ethically sound, and meaningful for both researchers and players. Through professional game developer insights, we highlight what it takes to develop a successful citizen science video game for cognitive science, and why this approach is still rare in the literature.

# Introduction

Citizen science is a scientific practice in which non-professional volunteers, or citizens, are engaged in the scientific process. It has a long history, dating back to the time of naturalists with the Audubon Society's Christmas Bird Count, established in 1900 (Dickinson et al., 2012). But citizen science is not limited to providing data; the volunteers may contribute in different ways such as co-designing protocols with the scientists (Haklay et al., 2021). With technological advances, the scale and sophistication of citizen science projects have expanded considerably; it is now a powerful tool in all scientific fields, for instance in ecological and environmental sciences (Fraisl et al., 2022), clinical science (Heyen et al., 2022), mathematics (Hartkopf 2019) or astronomy (Lintott et al., 2008).

More recently, cognitive scientists have begun to adopt similar approaches (Van den Bussche et al., 2024, 2026). A distinctive feature of cognitive science is that the participants' behavior is precisely the subject of the research. This specificity raises methodological and ethical considerations, particularly regarding experimental control, data reliability, and informed consent. At the same time, citizen science allows the study of cognition in naturalistic settings and across diverse populations, enhancing both the ecological validity and inclusivity of cognitive science research.

In this review, we will focus on the use of video games for cognitive citizen science. We will see that video games provide scientists with access to a population that would otherwise be unreachable, improving ecological validity, engagement, and data quality. We will also review the limitations of this approach, including the potential ethical issues and biases that could affect the reliability of the data collected. Finally, we will provide readers with practical advices from professional game developers on how to develop a successful video game for cognitive science research and avoid its inherent pitfalls.

# Video Games for Cognitive Citizen Science: a way forward.

Cognitive science, and more generally all research fields based on human data (clinical sciences, psychology, artificial intelligence…) has been hit by a wall: the replication crisis. In psychology for instance, the majority of published studies can't be replicated, which raises concerns on the credibility and validity of findings of the field (Open Science Collaboration, 2015). This crisis has several causes, including publication bias, researcher bias, and misuse of statistics. Several structural changes have been implemented by scientists to mitigate these issues. But arguably the most important cause of the crisis and the most difficult to solve is the low sample size and diversity of the populations used in cognitive science studies. Traditional laboratory studies in cognitive science often rely on small, homogeneous samples, typically undergraduate students, which restrict the validity and generalizability of the findings (Henrich et al., 2010).

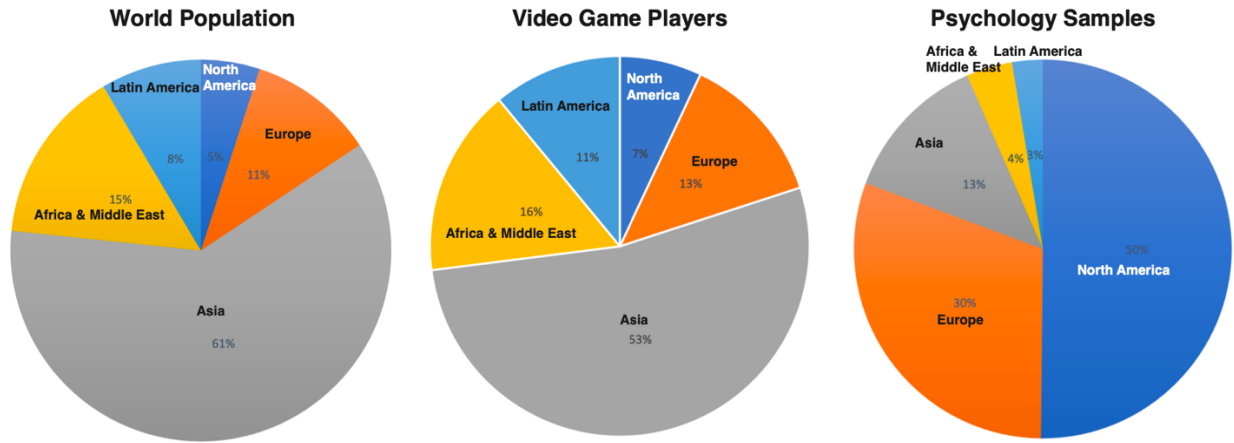

**Figure 1** - Distribution of three populations across the globe (from left to right): the World population (United Nations, 2024), video game players (Newzoo, 2025) and participants to psychology experiments. This last population has been derived from 311 samples of humans used in studies published between 2015 and 2019 in evolutionary psychology journals (Pollet & Saxton 2019).

It has been shown that 96% of the participants used in psychology studies only represent 12% of the planet (Arnet, 2008). Pollet & Saxton examined 311 samples of humans used in studies published between 2015 and 2019 in evolutionary psychology journals (Pollet & Saxton 2019). They classified 80% of the samples as "Western", the remaining coming from Asia (13%), and only a tiny fraction from Africa, the Middle East and Latin America (see the third pie chart in Figure 1).

**Increasing sample size and diversity**

Citizen science offers a mean to overcome these limitations by enabling large and demographically varied participant pools, including groups underrepresented in conventional research (Todowede et al., 2023). Cognitive scientists develop web, tablet, and mobile applications to collect ecologically valid data from diverse populations (Long et al., 2023; Allen et al., 2024). Video games like Skill Lab (Pedersen et al., 2023) and Sea Hero Quest (Spiers et al., 2023) exemplify how gamified cognitive research can leverage mobile platforms to gather rich data sets. Skill Lab, for instance, a mobile video game assessing diverse cognitive skills (e.g., working memory, reasoning) through engaging tasks, collected over 10,000 users' data. Similarly, Sea Hero Quest, a spatial orientation video game available on smartphones counted over 4 million players and helped better understand how humans navigate. Turning our – sometimes boring – tasks into video games has been proven to be a very fruitful approach. Indeed, humanity spends 24 billion hours a week playing video games (Newzoo, 2025). Diverting a fraction of this workforce to science would be a game changer. Video gaming has become a global phenomenon, with its popularity spreading across all demographics. We are far from the cliché of the male teenagers playing alone in their parent's basement. Instead, the average gamer in the US is now 35 years old and 45% female. In terms of geographical origin, the distribution of gamers closely matches that of the world population, see pie charts 1 and 3 in Figure 1. Although video gamers are not a perfect representation of the population, with some underrepresented demographics (see below), they are a significant improvement from our small in-lab samples, mainly made of college students. The Moral Machine, a game where players make moral decision, is a great example of diversity sampling, the platform gathered 40 million decisions in ten languages from millions of people in 233 countries and territories, showing cultural variation in moral judgments (Awad., 2018).

**Improving ecological validity**

In addition to improving scale and diversity, this approach enhances ecological validity. Laboratory environments, while allowing experimental control, often fail to capture the complexity of behavior in real-world contexts (Vigliocco et al., 2024). By collecting data from participants in their natural environments, on their smartphones, at home, or while commuting, citizen science projects provide valuable insights into how cognition operates in everyday life (Andrews et al., 2020; Pedersen et al., 2023). This approach bridges the gap between controlled experimentation and the realities of cognitive functioning in dynamic, uncontrolled settings (Velasco et al., 2026). CitieS-Health - NO2 and cognition in Barcelona (Gignac et al., 2022) is a good example. This project combines cognitive games and environmental data to study the impact of air pollution on brain functions, and identified a correlation between chronic $NO_2$ exposure and reduced cognitive performance.

**Improving data quality**

Beyond participation, gamification can improve data quality. Gamified tasks are associated with greater motivation, sustained attention, and cognitive effort among participants, which enhances the reliability of the data they provide (Alsawaier, 2018). For instance, inter-rater agreement has been shown to increase from poor to moderate levels when tasks are gamified (Poženel et al., 2022). This improvement is particularly valuable in remote and decentralized research contexts, where direct control over task compliance is limited (Rodd, 2024). On the other hand, tedious tasks, even when performed in a well-controlled lab environment, can lead to boredom and patterns of worsening behavioral performance with time-on-task (Zanesco et al., 2025). Mechanisms such as challenge, narrative framing, feedback, and social comparison help mitigate boredom and fatigue, maintaining engagement even in repetitive or demanding tasks (Genovese et al., 2024). The Great Brain Experiment validated this approach (Brown et al., 2014). Their collection of mini games explored individual differences in classic experimental psychology paradigms and showed that their large sample size (20,800 users) greatly reduced the noise inherent in data collection in an outside controlled laboratory setting and produced similar results to those of a comparable laboratory-based study.

**Improving engagement**

Gamification is commonly defined as "the use of game-design elements in non-game contexts" (Deterding et al., 2011). It offers substantial advantages for participant engagement, improving cognitive, motivational, and behavioural outcomes in different domains of application (Sailer et al., 2020). The integration of game elements, such as goals, feedback, progress tracking, and narrative framing into research tasks increases both the quantity and the reliability of collected data. Gamified tasks typically lead to higher participation and lower attrition rates. For instance, in mobile experience sampling studies, participants completing a gamified task submitted significantly more entries, including spontaneous, unprompted responses (Van Berkel et al., 2017). In large-scale market research, each additional gamified layer, such as interactive mechanics or narrative framing, was associated with higher task completion rates (Cechanowicz et al., 2013).

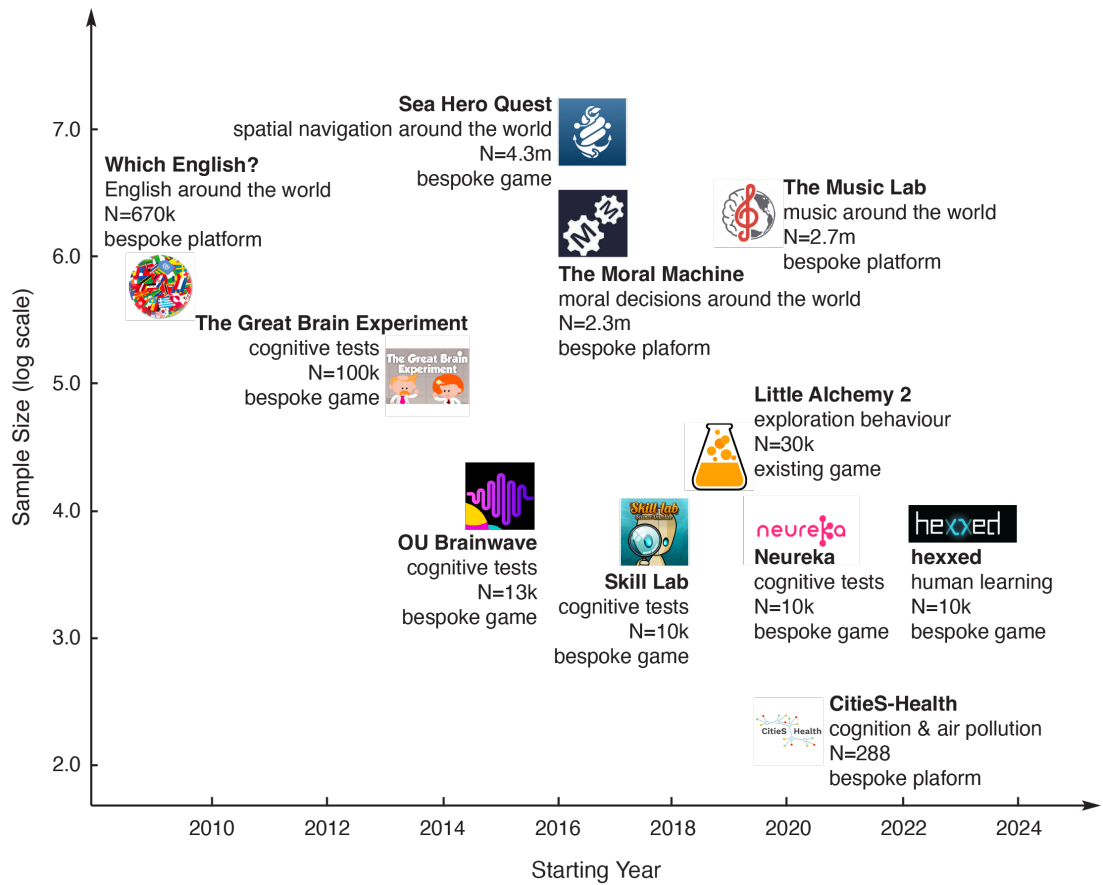

**Figure 2:** Cognitive Citizen Science Games : Sample Size as a function of starting year. For each game we show its logo, research area, sample size and whether it is a bespoke game or embedded in an existing game.

These effects appear consistent across age groups and levels of gaming experience. Results from cognitive studies can also be implemented back in video games to design better experiences that make repetitive or effortful tasks feel intrinsically rewarding. Work on enjoyment in games suggests that engagement peaks when challenges are calibrated to produce sustained learning progress—typically at intermediate difficulty—and when feedback helps players form accurate expectations and experience frequent, meaningful signals of success (Brändle et al., 2024). These principles can be translated into citizen-science task design via adaptive difficulty, clear progress cues, and reward structures that preserve scientific validity while reducing boredom and fatigue. For instance, Donegan et al. (2023) embedded the core structure of a cognitive planning task within Cannon Blast, a diamond-shooting smartphone game explicitly designed to be fun and repeatable, and deployed it through the Neureka app to collect data from participants at home, illustrating how game mechanics can maintain engagement across the many trials required for reliable behavioral measurement. But the most efficient way to create engagement is not to merely attract participants through superficial rewards, but to strengthen their intrinsic motivations and reinforce a sense of meaningful contribution, as detailed in the next section.

**Mutual Benefits for Researchers and Citizens**

The integration of gaming and citizen science generates reciprocal benefits for both researchers and participants. From a scientific perspective, citizen science games facilitate large-scale data collection, enhance statistical power, and improve the generalizability of findings, while substantially reducing the logistical and financial constraints associated with traditional laboratory research (Cooper et al., 2010). For instance, if Sea Hero Quest used a classic in-lab approach to gather its 4 million players dataset, it would have taken 500 years for a full-time PhD student (15 min/participant, 2000 hours per year) and have cost around 25 million euros in staff salary, 40 million euros in participants compensation (10€/participant). And this does not take into account the cost of flying participants from Japan, Brazil or South Africa to our French or British labs.

At the same time, these initiatives foster stronger connections between science and the public, narrowing the perceived divide between experts and non-experts. While many people express a willingness to contribute to scientific research, few have the time, knowledge, or motivation to take on formal roles. Video games offer an accessible and enjoyable way to get involved. Game developers aim at meeting players where they are, through game mechanics that are intuitive and emotionally resonant, while maintaining the scientific integrity of the tasks involved.

Participants also gain valuable experience from taking part in such projects. By embedding research within interactive and accessible formats, citizen science demystifies scientific concepts and processes, reducing apprehension toward participation (Speelman et al., 2023). Players often develop a deeper understanding of the scientific process, increasing their awareness and curiosity about the scientific process (Curtis, 2014). Some citizen science games, particularly those integrated into multiplayer environments, create communities of participants who collaborate toward shared goals, strengthening commitment and long-term engagement in science (Tinati et al., 2017; Waldispühl et al., 2020, Vergara-Borge et al., 2025). Collectively, these initiatives transform citizens from passive data providers into active contributors and co-creators of scientific knowledge. A great example is the Music Lab (Mehr et al., 2019): while giving valuable data for researchers on how music is created and perceive in humans, this app also gives participants knowledge about their own musical capacity and the cognitive process underlying it.

| Project | Research focus | Starting Year | Data collected | Tasks | Population | Sample Size | Game type | Developers | Funding | Communication / Motivation |
|---|---|---|---|---|---|---|---|---|---|---|
| **Sea Hero Quest** (Coutrot et al., 2018) | Spatial Cognition | 2016 | Spatial trajectories | Wayfinding + Path Integration | Global (all countries) | 4.3 M | Bespoke mobile app | Game developers | Industrial sponsorship | Large-scale "game for good" campaign |
| **Which English?** (Hartshorne et al., 2018) | English grammar variation | 2009 | Linguistic judgments | Rate grammaticality of English sentences | Native and non-native English speakers | 670,000 | Bespoke online platform | Academics | Gov + academic grants | Social media engagement |
| **The Music Lab** (Mehr et al., 2019) | Cross-cultural music cognition | 2019 | Beat, tuning, melody perception | Online musical tasks | Global | 2.7 M | Online platform | Academics | Gov + academic grants | Broad dissemination via website and media |
| **CitieS-Health** (Gignac et al., 2022) | Air pollution & cognition | 2020 | Cognitive & mental health tests, GPS | Stroop-like tasks | Barcelona residents | 288 | Bespoke mobile app | Academics | EU Horizon 2020 | Local campaigns, health orgs, social media |
| **Little Alchemy 2** (Brändle et al., 2023) | Exploration behavior & empowerment | 2019 | Gameplay logs | Combine elements to form complex objects | Global | 29,493 | Existing commercial game | Game developers | Academic + industry | Publication-driven outreach |
| **Skill Lab** (Pedersen et al., 2023) | Cognitive ability profiling | 2018 | Cognitive tests | Six mini-games + 14 validated tasks | Danish participants | 10,725 | Bespoke app | Game developers + Academics | ERC H2020 + industry | TV collaboration, school outreach |
| **Neureka** (Rosická et al., 2024) | Brain health & mental risk screening | 2020 | Cognitive tasks + demographics | Mini games on working memory & planning | Global (54 countries) | 9,918 | Bespoke app | Academics | Gov + academic grants | TV/radio ads, SciStarter, paid sub-studies |
| **The Great Brain Experiment** (Brown et al., 2014) | Smartphone-based cognitive phenotyping | 2013 | Cognitive + demographic data | Inhibition, visual & working memory | Global | 100,000 | Bespoke app | Game developers + Academics | Academic grants | Twitter, media blogs (e.g., WSJ Speakeasy) |
| **Moral Machine** (Awad., 2018) | Ethics of autonomous vehicles | 2016 | Moral decisions | Choose outcomes in moral dilemmas | Global (all countries) | 2.3 M | Bespoke online platform | Academics | Academic grants | Media coverage, TEDx talks, publications |
| **Hexxed.io** (Quendera et al., 2022) | human learning process | 2022 | Interaction logs (tap+swipe) | no instructions, find the patterns that maximize the number of points | Global | 10,000 | Bespoke app | Academics | Academic grant | Local campaigns, publications, social media |

Table 1: Taxonomy of cognitive citizen science video games.

## Creating a video game for cognitive citizen science: one goal, several recipes

Figure 3 proposes a pipeline summarizing the main steps to create a cognitive citizen science. The first stage involves clearly defining the research question and cognitive processes of interest and selecting appropriate task structures (single vs. multi-task). A key design decision is whether to follow a game-first model (focused on gameplay, e.g., Sea Hero Quest, Spiers et al., 2023) or a problem-first model (centered on scientific challenges with narrative elements, e.g., Foldit, Cooper et al., 2010). Once the model is chosen, development can proceed through partnerships with existing platforms, professional game studios, or internal teams. Interdisciplinary collaborations with fields like educational science or behavioral economics can provide valuable resources and expertise. Design quality and user experience are critical for participant engagement and data reliability. Internal development allows full experimental control but may lack polish. Finally, successful dissemination relies on strategic communication, leveraging media, gaming communities, and word-of-mouth, to recruit and retain a motivated player base over time. Hartshorne et al. (2018) used this strategy in Which English?, a linguistic game in which participants help to identify differences in English grammar around the world. The game had strong social media engagement, leading to the collection of data from over 670,000 players.

**Why Gamification Still Seems Risky to Researchers**

A few researchers have already tried working with developers and left the experience frustrated. The core issue is not simply technical failure, but a mismatch of goals and constraints. Developers often prioritize features that increase game fluidity, visual appeal, or user enjoyment, whereas researchers need tight control over variables, reliable data capture, and transparent task logic (Bunt et al., 2024). This divergence frequently leads to miscommunication, wasted effort, or gamified tools that look polished but compromise scientific rigor. Researchers come away with the sense that "gamification doesn't work for science", but the real problem is that the game was not built for science (Allen et al., 2024).

Another widespread belief is that gamified tools are too costly to justify, especially for exploratory or early-stage research. In academic circles, rough estimates for serious game development often range from €50,000 to €300,000 for a custom-built application (Bramble, 2023). For many labs, this is simply out of reach, leading to a preference for DIY solutions or traditional surveys.

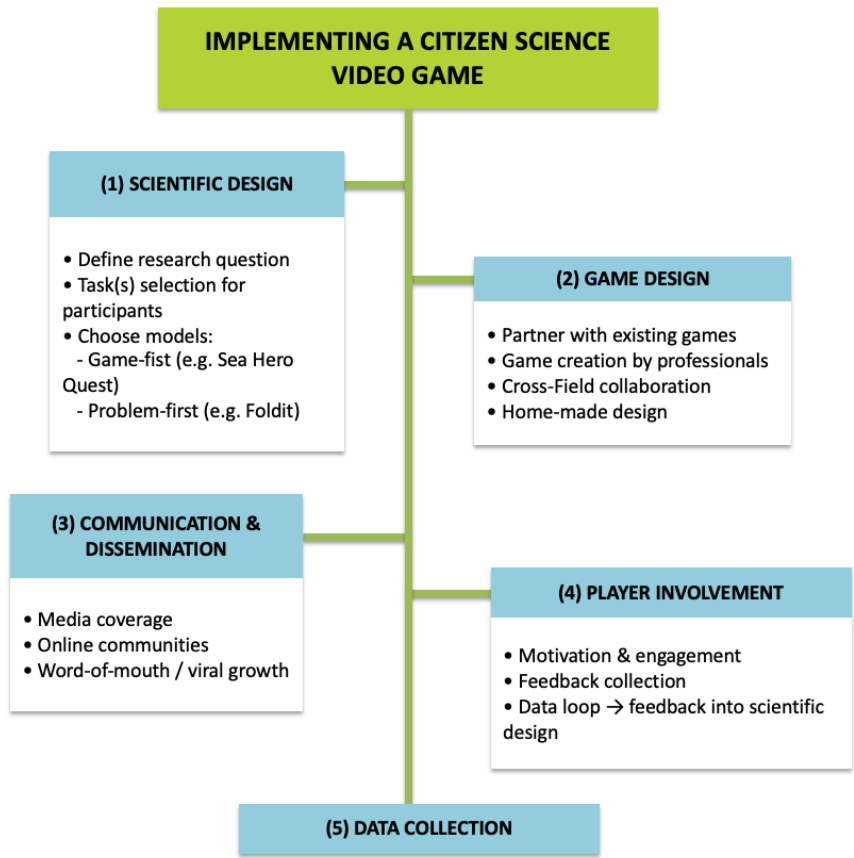

**Figure 3:** Implementation pipeline for Citizen Science games in Cognitive Science research

Many labs believe the solution is to internalize development, often by encouraging doctoral students or postdocs to build digital tools themselves. This strategy appears efficient, as it promotes the training of versatile researchers capable of both theoretical reasoning and computational implementation. But in practice, this model suffers from a lack of division of labor. Developing a high-quality gamified tool is not a side skill; it requires proficiency in UX, game design, front-end and back-end logic, data integrity, and testing. For early-career researchers, time spent learning or implementing these skills often comes at the expense of theoretical innovation, data analysis, or publishing. What looks like versatility can be a huge opportunity cost, especially given the rising availability of specialized teams who already know how to build scientifically robust gamified tools. As research grows more interdisciplinary, relying on dedicated collaborators, rather than overloading individual researchers, may be a more scalable and productive approach.

In summary, although many researchers are still reluctant to engage with gamification, this reluctance is due to past structural limitations rather than any intrinsic flaws in the method. With better communication models, more affordable tools, and smarter allocation of labor, these obstacles are increasingly surmountable (Long et al., 2023).

# Current limits

### Ethics

The growth of citizen science in cognitive research brings new challenges and raises key questions. Large-scale data collection often involves sensitive information, including clinical, demographic, or geographic details. Although researchers routinely obtain informed consent and anonymize data to protect participant privacy, ethical concerns about data security and transparency persist, especially when corporate entities are involved (Resnik, 2019). These issues highlight the need for clearer frameworks governing how data are stored, shared, and reused across citizen science platforms. This is particularly true when working with children, a public particularly receptive to video games, but whose informed consent is more complicated to obtain (Reiheld & Gay, 2019).

### Bias & accessibility

Another major concern relates to bias and accessibility. Despite the promise of broader representation, citizen science projects may still reproduce social and technological inequalities. For example, video games requiring access to smartphones or stable internet connections can unintentionally exclude certain demographics, while voluntary participation introduces self-selection bias, favoring individuals already interested in gaming or science (Stone et al., 2024). This self-selection bias remains a structural challenge: participants who engage with citizen science often possess higher digital literacy or motivation than the general population, potentially skewing results and limiting generalizability (Stone et al., 2024). This bias tends to strengthen with time: longitudinal audits reveal a persistent "long-tail" pattern, with only a fraction of participants sustaining activity beyond the first month, leading to the emergence of a core group of highly engaged contributors, while the majority contribute sporadically or disengage entirely (Fisher et al., 2021). Recent findings further complicate this picture by revealing cognitive and experiential disparities between frequent and infrequent gamers, even when both groups volunteer to participate. A study using the cognitive game *Tunnel Runner* found that infrequent gamers reported lower usability, reduced clarity of goals, and greater frustration compared to frequent gamers, despite comparable levels of focus and aesthetic appreciation (Markovitch, Markopoulos et al., 2024). Importantly, these experiential differences can translate into differences in data quality, as detailed in the next section.

Thus, inclusion in citizen science is not merely a question of access but also of equitable interaction. Designing for broader participation requires proactively identifying and mitigating barriers that certain user groups face, such as excessive cognitive load, unfamiliar mechanics, or unintuitive feedback systems. Infrequent gamers, for instance, could benefit from preparatory mini-games, clearer instructions, or participatory design methods that include them early in development. Otherwise, citizen science initiatives risk reinforcing the very participation gaps they aim to bridge (Markovitch, Markopoulos et al., 2024). In sum, while citizen science holds great potential for democratizing cognitive research, careful attention must be paid to experiential and performance asymmetries between participant subgroups to ensure inclusivity and data validity.

**Data quality and validation**
A frequent critique of citizen science is that variability in participant engagement and experience may lead to inconsistent or unreliable data (Cai, 2024). This is for instance the case in Tunnel Runner, mentioned above: while frequent and infrequent gamers performed equally well on simple reaction time tasks, infrequent gamers showed significantly lower precision and validity in more complex cognitive tasks, such as response-rule switching and inhibition (Markovitch et al., 2025) These results suggest that less experienced gamers, who likely represent a large proportion of the general public, may generate noisier or less reliable data in game-based paradigms, potentially introducing unnoticed biases into large-scale cognitive datasets.

To address this, validation strategies comparing citizen-generated data to established experimental benchmarks are crucial to ensure scientific robustness and replicability (Balázs, 2021). Brown et al. (2014) exemplified this approach by developing a smartphone app embedding four classic cognitive tasks; within one month, over 20,000 users contributed data that successfully replicated established cognitive effects, despite the lack of experimental control. Although challenges such as device variability, environmental distractions, and data loss were acknowledged, the authors emphasized that the scale and accessibility of smartphone-based research provide a powerful complement to laboratory studies and a unique opportunity to engage the public in cognitive science.

**Public involvement**

Another emerging discussion revolves around the degree of public participation in the scientific process. While citizen involvement is a cornerstone of the citizen science ethos, participation in cognitive research has so far been mostly passive: scientists collect behavioral and cognitive data from individuals without necessarily involving them in study design or interpretation (Van den Bussche, 2024). Future efforts should aim to include citizens' voices throughout all stages of research, from conception to dissemination, ensuring that projects are co-created rather than simply crowd sourced. This participatory approach is particularly crucial when research involves indigenous or marginalized communities, whose perspectives and epistemologies are often overlooked in traditional scientific frameworks.

**Conclusion**

Citizen science video games are particularly important for cognitive science, as they can contribute to mitigate one of the greatest threats faced by this field: sample sizes and homogeneous participant profiles. However, this approach is a double-edged sword, as it comes with its own problems, the most important being the participation bias, leading to specific participant profiles and lower data quality. Sea Hero Quest for instance, collected data from 4 million participants across the world, but only a couple of countries in Africa had enough players to be actually included in the analysis (Coutrot et al., 2018). To decrease these biases, we need to actively develop citizen science in less represented communities, for instance by creating coalitions that include shared governance of projects by people from the Global North and South (Pandya et al., 2012; Cooper et al., 2023; Carlen et al., 2024). Video games are a great tool for this, as they can be collaboratively designed, and then inexpensively deployed across the world. One problem is that both video game development and building an inclusive scientific network requires time, more than the 3 to 5 years generally given by academic institutions and grants. This calls for re-thinking academic tenure and promotion evaluations, placing greater value on non-traditional markers of academic success like collaborating with community members, conducting outreach, and working with unconventional professionals such as game developers.

**Conflicts of Interest**

Maxwell Scott-Slade is co-founder and Game Director at GLITCHERS Ltd., the commercial studio that developed Sea Hero Quest. GLITCHERS currently explores commercial opportunities for Sea Hero Quest as a screening tool for dementia and as a novel cognitive assessment.

# References

Allen, K., Brändle, F., Botvinick, M., Fan, J. E., Gershman, S. J., Gopnik, A., ... & Schulz, E. (2024). Using games to understand the mind. *Nature Human Behaviour*, *8*, 1035–1043.

Alsawaier, R. S. (2018). The effect of gamification on motivation and engagement. *The International Journal of Information and Learning Technology*, *35*(1), 56–79.

Atari, M., Xue, M., Park, P., Blasi, D., & Henrich, J. (2023). Which humans? PsyArXiv.

Awad, E., Dsouza, S., Kim, R., Schulz, J., Henrich, J., Shariff, A., ... & Rahwan, I. (2018). The Moral Machine experiment. *Nature*, *563*(7729), 59–64.

Balázs, B., Mooney, P., Nováková, E., Bastin, L., & Arsanjani, J. J. (2021). Data quality in citizen science. In K. Vohland, A. Land-Zandstra, L. Ceccaroni, R. Lemmens, J. Perelló, M. Pontier, T. Samson, & J. Wehn (Eds.), *The science of citizen science* (pp. 139–157). Springer.

Bramble, R. (2023). *How much does it cost to make a video game?* GameMaker. https://gamemaker.io/fr/blog/cost-of-making-a-game (retrieved March 2026)

Brändle, F., Stocks, L. J., Tenenbaum, J. B., Gershman, S. J., & Schulz, E. (2023). Empowerment contributes to exploration behaviour in a creative video game. *Nature Human Behaviour*, *7*(9), 1481–1489.

Brouwer, S., & Hessels, L. K. (2019). Increasing research impact with citizen science: The influence of recruitment methods on sample composition and data quality in pilot projects. *Public Understanding of Science*, *28*(5), 606–621.

Brown, H. R., Zeidman, P., Smittenaar, P., Adams, R. A., McNab, F., Rutledge, R. B., & Dolan, R. J. (2014). Crowdsourcing for cognitive science: The utility of smartphones. *PLOS ONE*, *9*(7), e100662.

Bunt, L., Greeff, J., & Taylor, E. (2024). Enhancing serious game design: Expert-reviewed, stakeholder-centered framework. *JMIR Serious Games*, *12*(1), e48099.

Carlen, E. J., Estien, C. O., Caspi, T., Perkins, D., Goldstein, B. R., Kreling, S. E., ... & Schell, C. J. (2024). A framework for contextualizing social-ecological biases in contributory science data. *People and Nature*, *6*(2), 377–390.

Carlier, S., Backere, F. D., & Turck, F. D. (2024). *Personalised serious games and gamification in healthcare: Survey and future research directions*. arXiv, 2411.18500.

Cechanowicz, J., Gutwin, C., Brownell, B., & Goodfellow, L. (2013). Effects of gamification on participation and data quality in a real-world market research domain. In *Proceedings of the First International Conference on Gameful Design, Research, and Applications* (pp. 58–65). ACM.

Cooper, S., Khatib, F., Treuille, A., Barbero, J., Lee, J., Beenen, M., ... & Players, F. (2010). Predicting protein structures with a multiplayer online game. *Nature*, *466*(7307), 756–760.

Cooper, C., Martin, V., Wilson, O., & Rasmussen, L. (2023). Equitable data governance models for the participatory sciences. *Community Science*, *2*(2), e2022CSJ000025.

Coutrot, A., Silva, R., Manley, E., De Cothi, W., Sami, S., Bohbot, V. D., Wiener, J. M., Hölscher, C., Dalton, R. C., Hornberger, M., & Spiers, H. J. (2018). Global Determinants of Navigation Ability. Current Biology, 28(17), 2861-2866.e4. https://doi.org/10.1016/j.cub.2018.06.009

De Cocker, K., Chastin, S. F., de Bourdeaudhuij, I., Imbo, I., Stragier, J., & Cardon, G. (2019). Citizen science to communicate about public health messages: The reach of a playful online survey on sitting time and physical activity. *Health Communication*, *34*(7), 720–725.

Curtis, V. (2014). Online citizen science games: Opportunities for the biological sciences. *Applied & Translational Genomics*, *3*(4), 90–94.

den Houting, J., Higgins, J., Isaacs, K., Mahony, J., & Pellicano, E. (2021). 'I'm not just a guinea pig': Academic and community perceptions of participatory autism research. *Autism*, *25*(2), 148–163.

Deterding, S., Dixon, D., Khaled, R., & Nacke, L. (2011). From game design elements to gamefulness: Defining "gamification." In *Proceedings of the 15th International Academic MindTrek Conference: Envisioning Future Media Environments* (pp. 9–15). ACM.

Dickinson, J. L., Zuckerberg, B., & Bonter, D. N. (2010). Citizen science as an ecological research tool: Challenges and benefits. *Annual Review of Ecology, Evolution, and Systematics*, *41*(1), 149–172.

Fischer, H., Cho, H., & Storksdieck, M. (2021). Going beyond hooked participants: The nibble-and-drop framework for classifying citizen science participation. *Citizen Science: Theory and Practice*, *6*(1).

Fraisl, D., Hager, G., Bedessem, B., Gold, M., Hsing, P.-Y., Danielsen, F., ... & Haklay, M. (2022). Citizen science in environmental and ecological sciences. *Nature Reviews Methods Primers*, *2*(1), 64.

Genovese, F., Bolognesi, M. M., Di Iorio, A., & Vitali, F. (2024). The advantages of gamification for collecting linguistic data: A case study using Word Ladders. *Online Journal of Communication and Media Technologies*, *14*(2), 1–17.

Gignac, F., Righi, V., Toran, R., Errandonea, L. P., Ortiz, R., Mijling, B., ... & Basagaña, X. (2022). Short-term NO2 exposure and cognitive and mental health: A panel study based on a citizen science project in Barcelona, Spain. *Environment International*, *164*, 107284.

Gillan, C. M., & Rutledge, R. B. (2021). Smartphones and the neuroscience of mental health. *Annual Review of Neuroscience*, *44*(1), 129–151.

Haklay, M., Dörler, D., Heigl, F., Manzoni, M., Hecker, S., & Vohland, K. (2021). What is citizen science? The challenges of definition. In K. Vohland, A. Land-Zandstra, L. Ceccaroni, R. Lemmens, J. Perelló, M. Pontier, T. Samson, & J. Wehn (Eds.), *The science of citizen science* (pp. 13–33). Springer.

Hartkopf, A. M. (2019). Developments towards mathematical citizen science. In *Forum Citizen Science 2019*.

Hartshorne, J. K., Tenenbaum, J. B., & Pinker, S. (2018). A critical period for second language acquisition: Evidence from 2/3 million English speakers. *Cognition*, *177*, 263-277.

Henrich, J., Heine, S. J., & Norenzayan, A. (2010). The weirdest people in the world? *Behavioral and Brain Sciences*, *33*(2–3), 61–83.

Heyen, N. B., Gardecki, J., Eidt-Koch, D., Schlangen, M., Pauly, S., Eickmeier, O., ... & Bratan, T. (2022). Patient science: Citizen science involving chronically ill people as co-researchers. *Journal of Participatory Research Methods*, *3*(1).

Kosmala, M., Wiggins, A., Swanson, A., & Simmons, B. (2016). Assessing data quality in citizen science. *Frontiers in Ecology and the Environment*, *14*(10), 551–560.

Lintott, C. J., Schawinski, K., Slosar, A., Land, K., Bamford, S., Thomas, D., ... & Vandenberg, J. (2008). Galaxy Zoo: Morphologies derived from visual inspection of galaxies from the Sloan Digital Sky Survey. *Monthly Notices of the Royal Astronomical Society*, *389*(3), 1179–1189.

Long, B., Simson, J., Buxó-Lugo, A., Watson, D. G., & Mehr, S. A. (2023). How games can make behavioural science better. *Nature*, *613*(7944), 433–436.

Markovitch, B., Kamps, J. C. C., Markopoulos, P., & Birk, M. V. (2025). Do cognitive assessment games leave infrequent video game players behind? Evaluating frequent and infrequent players' gaming experience and data quality. *Computers in Human Behavior*, *172*, 108720.

Mehr, S. A., Singh, M., Knox, D., Ketter, D. M., Pickens-Jones, D., Atwood, S., ... & Glowacki, L. (2019). Universality and diversity in human song. *Science*, *366*(6468), eaax0868.

Newzoo. (2025). *Global games market report 2025*. https://newzoo.com/resources/trend-reports/newzoo-global-games-market-report-2025 (retrieved March 2026)

Nugent, J. (2021). Citizen science: Accelerating Alzheimer's research with Stall Catchers. *The Science Teacher*, *88*(4), 16–17.

Pandya, R. E. (2012). A framework for engaging diverse communities in citizen science in the US. *Frontiers in Ecology and the Environment*, *10*(6), 314–317.

Pedersen, M. K., Díaz, C. M. C., Wang, Q. J., Alba-Marrugo, M. A., Amidi, A., Basaiawmoit, R. V., ... & Sherson, J. F. (2023). Measuring cognitive abilities in the wild: Validating a population-scale game-based cognitive assessment. *Cognitive Science*, *47*(6), e13308.

Poženel, M., Zrnec, A., & Lavbič, D. (2022). Measuring how motivation affects information quality assessment: A gamification approach. *PLOS ONE*, *17*(10), e0274811.

Quendera, T., Deng, D., Hamidi, M., Bergomi, M., Mainen, Z. F., & Agarwal, G. (2022). Humans learning a complex task are picky and sticky. In *Proceedings of the 2022 Conference on Cognitive Computational Neuroscience* (pp. 670–673).

Reiheld, A., & Gay, P. L. (2019). *Coercion, consent, and participation in citizen science*. arXiv, 1907.13061.

Resnik, D. B. (2019). Citizen scientists as human subjects: Ethical issues. *Citizen Science: Theory and Practice*, *4*(1), 11.

Rodd, J. M. (2024). Moving experimental psychology online: How to obtain high quality data when we can't see our participants. *Journal of Memory and Language*, *134*, 104472.

Rosická, A. M., Teckentrup, V., Fittipaldi, S., Ibanez, A., Pringle, A., Gallagher, E., ... & Gillan, C. M. (2024). Modifiable dementia risk factors associated with objective and subjective cognition. *Alzheimer's & Dementia*, *20*(11), 7437–7452.

Rowbotham, S., McKinnon, M., Leach, J., Lamberts, R., & Hawe, P. (2017). Does citizen science have the capacity to transform population health science? *Critical Public Health*, *29*(1), 118–128.

Sailer, M., & Homner, L. (2020). The gamification of learning: A meta-analysis. *Educational Psychology Review*, *32*, 77–112.

Spiers, H. J., Coutrot, A., & Hornberger, M. (2023). Explaining world-wide variation in navigation ability from millions of people: Citizen science project Sea Hero Quest. *Topics in Cognitive Science*, *15*(1), 120–138.

Speelman, E. N., Escano, E., Marcos, D., & Becu, N. (2023). Serious games and citizen science: From parallel pathways to greater synergies. *Current Opinion in Environmental Sustainability*, *64*, 101320.

Tinati, R., Luczak-Roesch, M., Simperl, E., & Hall, W. (2017). An investigation of player motivations in Eyewire, a gamified citizen science project. *Computers in Human Behavior*, *73*, 527–540.

Todowede, O., Lewandowski, F., Kotera, Y., Ashmore, A., Rennick-Egglestone, S., Boyd, D., ... & Slade, M. (2023). Best practice guidelines for citizen science in mental health research: Systematic review and evidence synthesis. *Frontiers in Psychiatry*, *14*, 1175311.

United Nations Department of Economic and Social Affairs, Population Division. (2024). *World population prospects 2024*. https://population.un.org/wpp/ (retrieved March 2026)

Van den Bussche, E., Verhaegen, K. A., Hughes, G., & Reynvoet, B. (2024). Towards a cognitive citizen science. *Nature Reviews Psychology*, *3*, 1–2.

Van den Bussche, E., Verhaegen, K. A., Reynvoet, B., & Hughes, G. (2026). Citizen science in psychology: Challenges, opportunities, and recommendations. *Behavior Research Methods*, *58*(5), 139.

Velasco, P. F., Coutrot, A., & Spiers, H. J. (2026). Human Navigation Behaviour and Brain Dynamics in Real-world Contexts. *arXiv preprint arXiv:2603.11347*.

Vergara-Borge, F., López-de-Ipiña, D., Emaldi, M., Olivares-Rodríguez, C., Khan, Z., & Soomro, K. (2025). Gamifying engagement in spatial crowdsourcing: An exploratory mixed-methods study on gamification impact among university students. *Systems*, *13*(7), 519.

Vigliocco, G., Convertino, L., De Felice, S., Gregorians, L., Kewenig, V., Mueller, M. A., ... & Spiers, H. J. (2024). Ecological brain: Reframing the study of human behaviour and cognition. *Royal Society Open Science*, *11*(11), 240762.

Vohland, K., Land-Zandstra, A., Ceccaroni, L., Lemmens, R., Perelló, J., Pontier, M., ... & Wehn, U. (Eds.). (2021). *The science of citizen science*. Springer Nature.

Waldispühl, J., Szantner, A., Knight, R., Caisse, S., & Pitchford, R. (2020). Leveling up citizen science. *Nature Biotechnology*, *38*(10), 1124–1126.

Yurdum, L., Singh, M., Glowacki, L., Vardy, T., Atkinson, Q. D., Hilton, C. B., ... & Mehr, S. A. (2023). Universal interpretations of vocal music. *Proceedings of the National Academy of Sciences*, *120*(37), e2218593120.

Zanesco, A. P., Denkova, E., & Jha, A. P. (2025). Mind-wandering increases in frequency over time during task performance: An individual-participant meta-analytic review. *Psychological Bulletin*, *151*(2), 217.